\documentclass[notitlepage,twocolumn]{revtex4-1}
\usepackage{mathtools}
\usepackage{bm}
\usepackage{amsmath,amssymb}
\usepackage{color}
\usepackage{braket}
\usepackage{graphicx}
\usepackage{epstopdf}
\usepackage{float}
\usepackage{nccmath}
\usepackage[normalem]{ulem} 
\usepackage{setspace}

\begin{document}
\title{Reply to ``Comment on `Winding around non-Hermitian singularities' by
	Zhong et al., Nat. Commun. 9, 4808 (2018)"}


\author{Q. Zhong$^1$}
\email[]{qizhong@mtu.edu}
\author{M. Khajavikhan$^2$}
\author{D.N. Christodoulides$^2$}
\author{R. El-Ganainy$^{1,3}$ }
\email[]{ganainy@mtu.edu}

\affiliation{$^1$ Department of Physics, Michigan Technological University, Houghton, Michigan, 49931, USA}
\affiliation{$^2$ College of Optics $\&$ Photonics-CREOL, University of Central Florida, Orlando, Florida, 32816, USA}
\affiliation{$^3$ Center for Quantum Phenomena, Michigan Technological University, Houghton, Michigan, 49931, USA}

\maketitle

A comment has been recently posted on the arXiv \cite{Pap2019arXiv} that discuss our recent work on encircling multiple exceptional points \cite{Zhong2018NC}. In that comment, the authors claim that our approach is prone to errors. In discussing their findings, they also indicate that a method presented by their team \cite{Pap2018PhysRevA} (presumed to give correct results in all situations) was published prior to our work. First, we would like to note that both their work and ours were posted on the arXiv within a week from each other. Second and more importantly, as we will show below, their analysis and conclusion concerning the validity of our approach as presented in the comment article \cite{Pap2019arXiv} are not correct. As we will demonstrate, a proper application of our method does indeed provide the correct results.

\begin{figure}[!t]
	\includegraphics[width=2.4in]{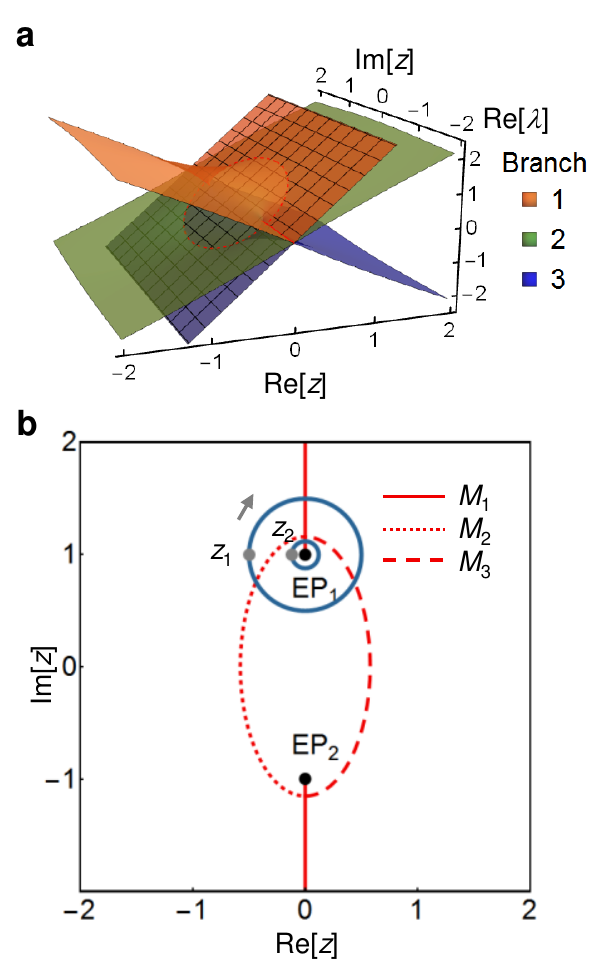}
	\caption{Real parts of the Riemann sheets associated with the eigenvalue solutions of the Hamiltonian $H$ of Eq. (\ref{H}) are depicted in (a). At every point in the complex domain $z$, the surfaces are sorted by their real values and colored accordingly. For clarity, we have indicated the solution $2z$ by a plain sheet. (b) The corresponding branch lines in the complex plane along with their associated permutation matrices $M_{1,2,3}$. Also we show two test loops encircling $\text{EP}_1$, which we discuss in details in the text.}
	\label{FigBranches}
\end{figure}

Let us first recall the example studied in \cite{Pap2019arXiv}:
\begin{equation}\label{H}
\begin{split}
H&=\begin{bmatrix}
1 & z & 0\\
z & -1 & 0\\
0 & 0 & 2z 
\end{bmatrix},
\end{split}
\end{equation}
which has the eigenvalues $\pm \sqrt{1+z^2}$ and $2z$. In our  approach, one would first pick a sorting scheme based on some chosen criterion. For example, we can sort the eigenvalues based on their magnitude, real part or imaginary part. After that, the eigenvalues must be sorted at every point in the complex plane according to the chosen method. This will naturally lead to a set of branch lines that separate the different solutions based on the sorting scheme. One can then associate a permutation matrix with each line to describe the transition between the different solution branches. We refer the interested reader to \cite{Zhong2018NC} for the detailed description of that procedure. Let us now apply this approach here. In order to make our point, we follow \cite{Pap2019arXiv} and sort the eigenvalues based on their real part. This results in the coloring scheme for the Riemann surfaces shown in Fig. \ref{FigBranches}(a). By projecting this on the complex domain, we find that we have two branch points but four branch lines (see Fig. \ref{FigBranches}(b)). In contrast, the analysis in \cite{Pap2019arXiv} identifies only two branch lines which is inconsistent with the sorting scheme. Back to Fig. \ref{FigBranches}(b), we note that the additional two lines (described by $M_2$ and $M_3$) are rather unusual branches since they do not end at branch points. This a direct outcome of the artificial nature of the example given in \cite{Pap2019arXiv} which does not arise any realistic physical situations (it amounts for analyzing two separate, uncoupled experiment by grouping their data together). Next, by inspecting the connectivity of the sheets across these branch lines, we can identify the following permutation matrices: 

\begin{equation}\label{Matrix}
\renewcommand\arraystretch{1}
\begin{split}
M_1&=\begin{bmatrix}
0 & 0 & 1\\
0 & 1 & 0\\
1 & 0 & 0 
\end{bmatrix},~ 
M_2=\begin{bmatrix}
1 & 0 & 0\\
0 & 0 & 1\\
0 & 1 & 0  
\end{bmatrix},~ 
M_3=\begin{bmatrix}
0 & 1 & 0\\
1 & 0 & 0\\
0 & 0 & 1 
\end{bmatrix}. 
\end{split}
\end{equation}

We can now use the above matrices to study the stroboscopic evolution of the eigenstates around any arbitrary loop. For illustration purpose, we consider the two loops shown in Fig. \ref{FigBranches}(b). The starting point for the larger loop is denoted by $z_1$. The loop crosses the lines associated with the matrices $M_{1,3,2}$ in that order. Thus after one cycle, the exchange relation of the states is given by the matrix product  $ M_2 M_3 M_1 (s_1, s_2, s_3)^\mathsf{T}=(s_2, s_1, s_3)^\mathsf{T}$. In simple terms, this means that state $s_1$ associated with the solution $\sqrt{1+z^2}$  will swap with state $s_2$ associated with $-\sqrt{1+z^2}$, as expected. On the other hand, state $s_3$ which is associated with the solution $2z$ remains on the same sheet, also as expected. Moving to the smaller loop with the starting point $z_2$, we note that that it crosses only the line associated with $M_1$ which results in  $ M_1 (s_1, s_2, s_3)^\mathsf{T}=(s_3, s_2, s_1)^\mathsf{T}$. These are the expected results since at point $z_2$, the states $s_{1,2,3}$ belong to the solutions $\sqrt{1+z^2}$, $2z$, $-\sqrt{1+z^2}$, respectively (because $\text{Re}[\sqrt{1+z_2^2}]>\text{Re} [2z_2]>\text{Re}[-\sqrt{1+z_2^2}]$).   

Interestingly, the authors in \cite{Pap2019arXiv}, make a comment also about the efficiency of various approaches. It is not clear to us how can one compare the efficiency of various methods without having a rigorous mathematical definition for the term `efficiency'.     

In summary, we have addressed the comments raised in \cite{Pap2019arXiv} and have shown that its conclusion is wrong.To illustrate this, we used the same example put forward in \cite{Pap2019arXiv} and demonstrated that our method provides indeed the correct results.

\bibliography{Reference}

\end{document}